\documentclass[]{spie}  

\usepackage{amsmath,amsfonts,amssymb}
\usepackage{graphicx}
\usepackage[colorlinks=true, allcolors=blue]{hyperref}
\usepackage{float}
\usepackage{caption}
\usepackage{subcaption}

\usepackage[english]{babel}
\usepackage{csquotes}
\usepackage[backend=biber]{biblatex}
\title{Estimating effective wind speed from Gemini Planet Imager's adaptive optics data using covariance maps}

\author[a]{Daniel M. Levinstein}
\author[a]{Saavidra Perera}
\author[a]{Quinn M. Konopacky}
\author[b]{Alex Madurowicz}
\author[b]{Bruce Macintosh}
\author[c]{Lisa Poyneer}
\author[d]{Richard W. Wilson}
\affil[a]{University of California San Diego, 9500 Gilman Drive, La Jolla, CA}
\affil[b]{Stanford University, 450 Serra Mall, Stanford, CA}
\affil[c]{Lawrence Livermore National Lab, 7000 East Ave, Livermore, CA 94550}
\affil[d]{Center for Advanced Instrumentation, University of Durham, South Rd, Durham DH1 3LS, United Kingdom}
\authorinfo{Send correspondence to D. M. Levinstein. E-mail: dmlevinstein@ucsd.edu}

\begin{document} 
\maketitle

\begin{abstract}
The Earth’s turbulent atmosphere results in speckled and blurred images of astronomical objects when observed by ground based visible and near-infrared telescopes. Adaptive optics (AO) systems are employed to reduce these atmospheric effects by using wavefront sensors (WFS) and deformable mirrors. Some AO systems are not fast enough to correct for strong, fast, high turbulence wind layers leading to the wind butterfly effect, or wind-driven halo, reducing contrast capabilities in coronagraphic images. Estimating the effective wind speed of the atmosphere allows us to calculate the atmospheric coherence time. This is not only an important parameter to understand for site characterization but could be used to help remove the wind butterfly in post processing. Here we present a method for estimating the atmospheric effective wind speed from spatio-temporal covariance maps generated from pseudo open-loop (POL) WFS data. POL WFS data is used as it aims to reconstruct the full wavefront information when operating in closed-loop. The covariance maps show how different atmospheric turbulent layers traverse the telescope. Our method successfully recovered the effective wind speed from simulated WFS data generated with the soapy python library. The simulated atmospheric turbulence profiles consist of two turbulent layers of ranging strengths and velocities. The method has also been applied to Gemini Planet Imager (GPI) AO WFS data. This gives insight into how the effective wind speed can affect the wind-driven halo seen in the AO image point spread function. In this paper, we will present results from simulated and GPI WFS data.
\end{abstract}


\section{INTRODUCTION}
\label{sec:intro}  

As light from distant stars passes through the atmosphere it becomes distorted resulting in blurred and speckled images.  Adaptive optics (AO) systems are employed in ground based telescopes to correct for wavefront aberrations caused by atmospheric turbulence.  However, in the case of high contrast imaging, if the coherence time is too low, the AO system cannot correct for these aberrations fast enough. Coherence time is a measure of the timescale of atmospheric turbulence variation, defined as
\begin{equation}
    \tau_0 = 0.314\frac{r_0}{v_{\text{eff}}},
    \label{eqn:coherence}
\end{equation}
where $r_0$ is the Fried parameter and $v_{\text{eff}}$ is defined as 

\begin{equation}
    v_{\text{eff}} = \left[ \frac{\int_0^\infty C_n^2(h)V(h)^\frac{5}{3}dh}{\int_0^\infty C_n^2(h) dh} \right]^\frac{3}{5},
    \label{eqn:veff integral}
\end{equation}
where $C_n^2(h)$ is the refractive index structure constant used to quantify the optical turbulence strength of layers at height h and $V(h)$ is the velocity profile at a given height \cite{Hardy98}. 
Low coherence time can result in the wind butterfly effect, or wind-driven halo, appearing in coronagraphic images \cite{Madurowicz18,Cantalloube20}. The wind butterfly effect, in which two lobes reminiscent of a butterfly's wings overlay the image, reduces contrast capabilities in coronagraphic images and is seen in images from GPI \cite{GPIES}. In this paper, we discuss a method of measuring the effective wind speed of the atmosphere using covariance maps constructed using GPI WFS data sets \cite{GPITelem}.




The methods section of this paper will describe the process of calculating the effective wind speed from simulated WFS slopes data generated using the soapy python library. Following this, results from simulated WFS data as well as GPI WFS data will be presented.

\section{METHODS}
As described by equations \ref{eqn:coherence} and \ref{eqn:veff integral}, knowledge of the atmospheric wind profile i.e. the wind speed and turbulence strength of each turbulent layer, is required to estimate the coherence time. These layers can be identified and characterized by calculating the spatio-temporal covariance maps of open-loop WFS slopes. A covariance map gives the joint variability of every pair of subapertures, for a given separation, in an AO WFS dataset at a given temporal offset ($\delta t$). This is defined as 
\begin{equation}
        A_{\delta i,\delta j, \delta t} = \langle C_{i,j,t}C'_{i',j',t'}\rangle,
    \label{eqn:covariance}
\end{equation}
\noindent where $C$ and $C'$ are the WFS slopes at subaperture position indices [$i,j$] and [$i',j'$] at times $t$ and $t'$, and [$\delta i,\delta j$] are the positional subaperture separations \cite{Wilson02,PREX}. With no temporal offset, i.e. $\delta t=0$, all wind layers are superimposed at the center of the covariance map. Assuming Taylor's frozen flow hypothesis, introducing a temporal offset causes the different turbulent layers to separate and traverse the map revealing their wind speed and relative strength. For a given $\delta t$ this creates a wind speed map that describes the speed at every location on the covariance map. Summing up the velocities of every peak weighted by the relative strength of the peak gives the effective wind speed of the atmosphere. We use an approximate equation for $v_{\text{eff}}$, defined as 
\begin{equation}
    v_{\text{eff}}^{\frac{5}{3}} = \alpha_1 v_1^{\frac{5}{3}} + ... + \alpha_n v_n^{\frac{5}{3}},
    \label{eqn:veff}
\end{equation}

\noindent where $v_n$ is the speed at a given subaperture separation and $\alpha_n$ is the relative strength. We can use this approximation because $\tau_0$ is only dependent on the effective wind speed and does not require an altitude profile.  As shown in equation \ref{eqn:coherence}, $r_{0}$ is also required to calculate $\tau_0$. $r_0$ can be calculated from the covariance map as well, as seen in Wilson 2002 \cite{Wilson02}. However, this calculation is not addressed in this paper.

Open-loop slopes are necessary to make a covariance map. However, since concurrent open-loop slopes and coronagraphic images are not possible, we calculate POL slopes using the following equation

\begin{equation}
    s_n^{POL} = s_n^{RES} + P \cdot a_{n-1},
\end{equation}

\noindent where $s$ is the slope, $P$ is the interaction matrix, and $a$ is the actuator demands from the previous frame (n-1) \cite{POL}.

The soapy python library was used to generate simulated Shack-Hartmann POL WFS slopes under a range of atmospheric conditions \cite{Soapy}. The description of the simulated AO system is described in Table \ref{tab:soapy}. The telescope diameter was based on the Gemini South Telescope and the subapertures were determined by maximizing the resolution while minimizing calculation time.

\begin{table}[H]
    \caption{Soapy setup parameters}
    \label{tab:soapy}
    \begin{center}
        \begin{tabular}{|l|l|}
            \hline 
            Parameter & Value \\
            \hline
            Telescope Diameter & 8.1 m \\
            \hline
            Obscuration Diameter & 1.0 m \\
            \hline
            Lenselet Array & $16\times16$ \\
            \hline
            Framerate & 1 kHz \\
            \hline
            DM Actuators & 9 \\
            \hline
        \end{tabular}
    \end{center}
\end{table}

\begin{figure}[H]
\begin{minipage}[t]{0.45\textwidth}
    \centering
    \vspace{0pt}
    \includegraphics[width=3in]{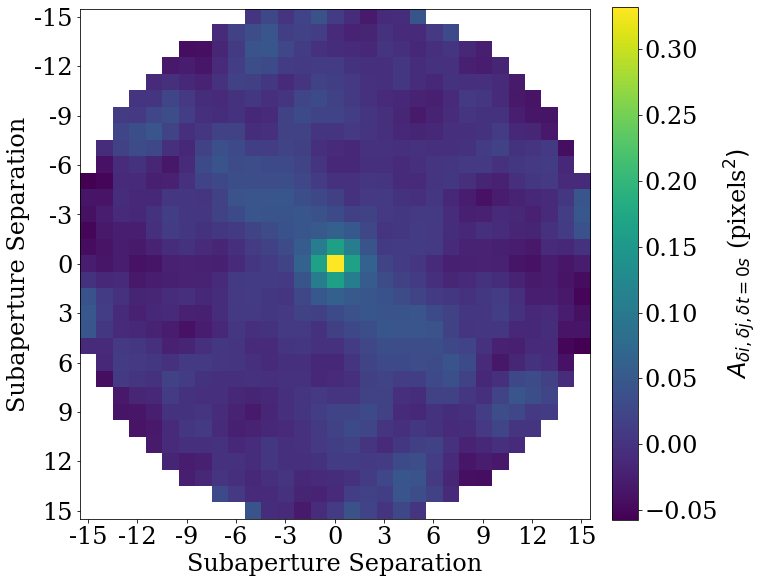}
    \caption{Covariance map with 0 temporal offset. Each subaperture slope has the highest covariance with itself, so there is a peak in the center of the image.}
    \label{fig:covmap_example}
\end{minipage}\hfill
\begin{minipage}[t]{0.45\textwidth}
    \centering    
    \vspace{0pt}
    \includegraphics[width=3in]{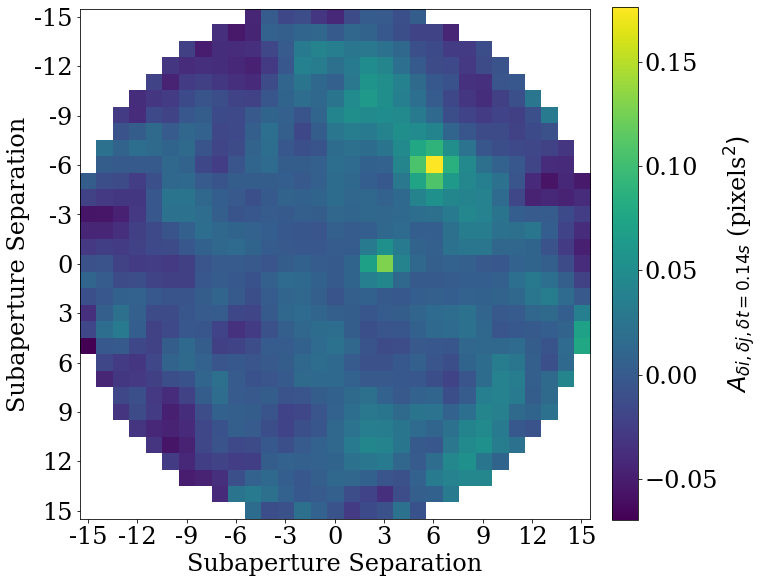}
    \caption{Covariance map with 0.14 s temporal offset. The two peaks represent two distinct turbulent layers.}
    \label{fig:temporal_covmap_example}
\end{minipage}\hfill
\end{figure}

Figures \ref{fig:covmap_example} and \ref{fig:temporal_covmap_example} are examples of covariance maps calculated using equation \ref{eqn:covariance} on simulated data. Without a temporal offset, each subaperture slope has the highest covariance with itself so all turbulent layers are superimposed at the central peak as seen in figure \ref{fig:covmap_example}. Introducing a temporal offset of $\delta t =  0.14$ s (140 frames with a framerate of 1000 Hz) yields the covariance map in figure \ref{fig:temporal_covmap_example}. The offset causes the superimposed wind layers to traverse the covariance map and separate into distinct wind layer peaks. In this example, there is a slow moving wind layer at 0 degrees (moving left to right) from the center and a faster moving wind layer at 45 degrees from the center.

\begin{figure}[H]
    \includegraphics[width=3in]{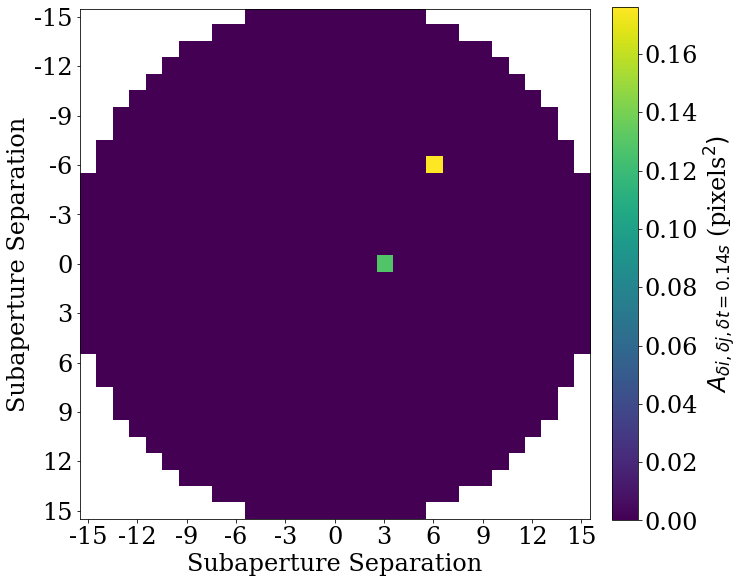}
    \centering
    \caption{Covariance map from figure \ref{fig:temporal_covmap_example} with power-normalization mask with threshold of 0.2. This process isolates the wind layers and removes the noise.}
    \label{fig:pow_norm_thresh_covmap}
\end{figure}

Adding up the speeds of the peaks weighted by their relative strengths according to equation \ref{eqn:veff} yields the effective wind speed of the atmosphere \cite{Saavi18}. A mask is made from a power normalization and threshold applied to the original covariance map to remove noise and only count the speeds from the relevant peaks. Figure \ref{fig:pow_norm_thresh_covmap} shows an example of this mask applied to the covariance map from figure \ref{fig:temporal_covmap_example}. Using this masked covariance map with the corresponding speed map and equation \ref{eqn:veff} we get the effective wind speed of the atmosphere to be 23.3 m/s $\pm$ 1.6 m/s from an input effective velocity of 21.6 m/s.

The simulated slopes dimensions were chosen to be $16\times16$ to strike a good balance between resolution and CPU time to run the simulations and calculate the covariance maps. The GPI WFS slopes are $48\times48$, so they were resized via averaging to $16\times16$ in order to work with the model. This resize makes it as if GPI's AO system only had $16\times16$ subapertures, each with a side length 3 times longer. Though this reduces the resolution, most of the information is still present and the calculation time is significantly reduced.

\section{RESULTS}
\subsection{Simulated Results}

The method described in the last section was verified with two turbulent layers at different strengths and wind speeds. The strength ratios of the ground layer to the higher altitude layer were 50:50, 60:40, and 70:30. The speeds varied from 5-20 m/s at the ground layer and 10-40 m/s at the high altitude layer. The Gemini telescope does not operate in winds above 20 m/s so no GPI datasets will have any ground wind layers above 20 m/s. 

\begin{figure}[H]
    \centering
    \includegraphics[height=3in]{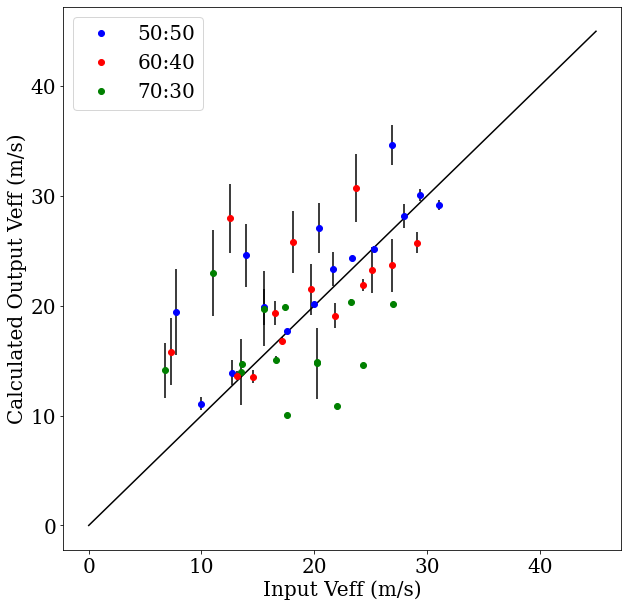}
    \caption{Measured effective wind speed vs input effective wind speed of simulated data. The colors of the points are related to the strength ratios: 50:50 is blue, 60:40 is red, 70:30 is green.}
    \label{fig:veff_correlation}
\end{figure}

Figure \ref{fig:veff_correlation} shows the correlation of the effective wind speed calculated from the covariance maps and the input effective wind speed. The correlation is more consistent for simulations that have each layer at the same relative strength. At strength ratios 60:40 and 70:30, fast, weak layers may not be strong enough to be distinguishable above the noise threshold as there is higher noise toward the edge of the covariance map. This causes the slower, stronger layer to be over represented in the effective wind speed calculation. The error bars are the calculated standard error for 10 repetitions of the same setup parameters.

\subsection{GPI Data Results}
\begin{figure}[H]
    \centering
    \includegraphics[height=3in]{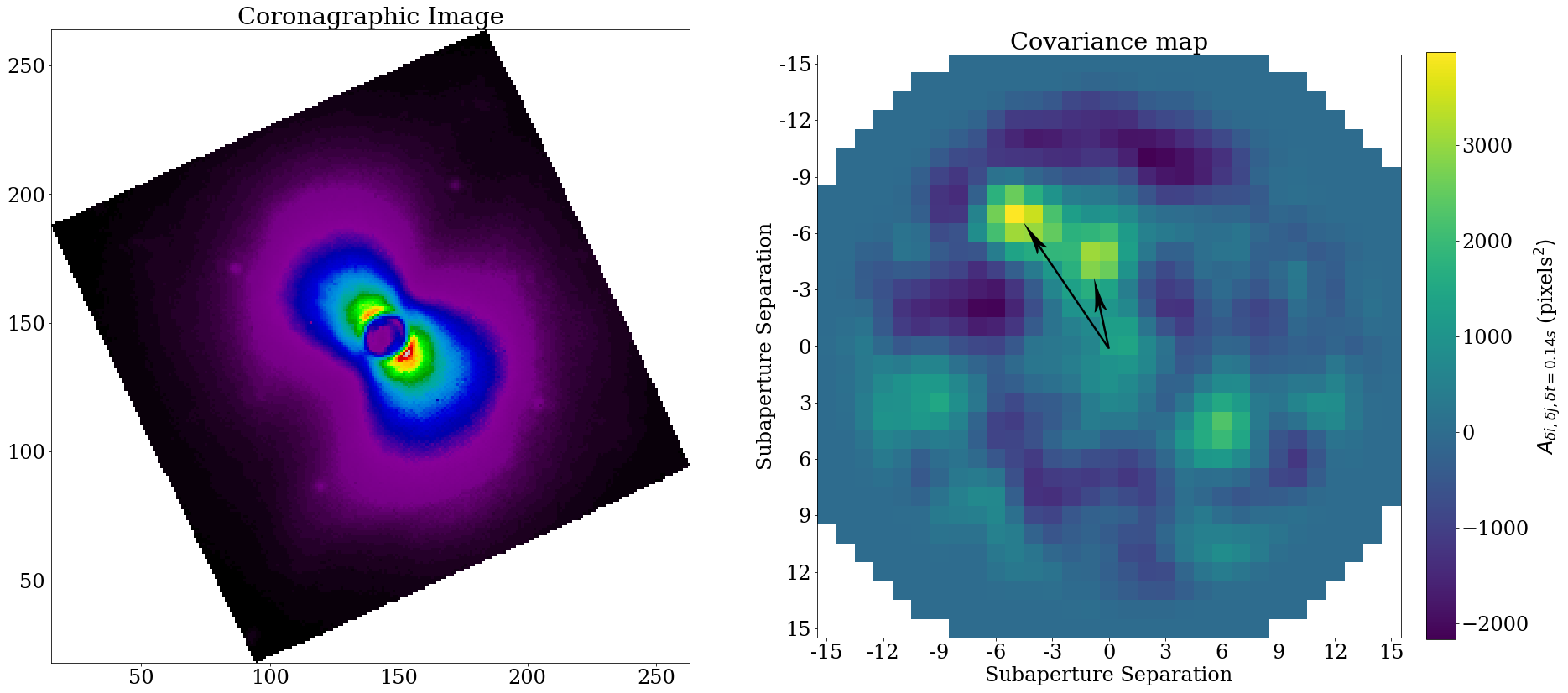}
    \caption{The coronagraphic image on the left has been weighted to enhance detail. The wind butterfly effect is very strong in this image. The associated covariance map with a temporal offset of 0.14 s on the right shows a strong fast wind layer and a weaker slower wind layer both moving up and left from the origin.}
    \label{fig:GPI}
\end{figure}

\begin{figure}[H]
    \centering
    \includegraphics[height=3in]{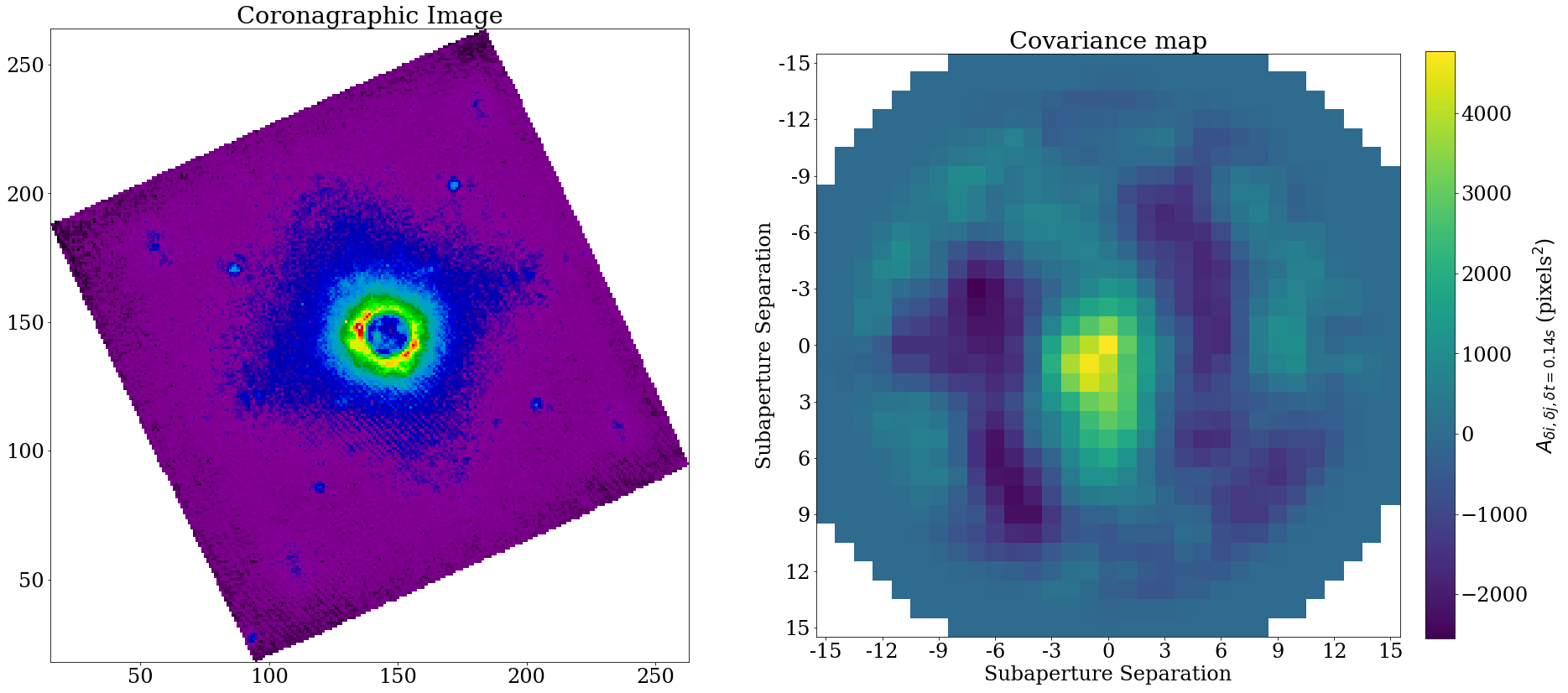}
    \caption{The wind butterfly effect is weak in this coronagraphic image. The associated covariance map on the right (0.14 s temporal offset) shows a strong slow wind layer moving down and left from the origin. It is unclear whether this is multiple turbulent layers or a single layer.}
    \label{fig:GPI_nofly}
\end{figure}

The method was applied to GPI POL WFS data to compare the effective wind speed to coronagraphic images that exhibit the wind butterfly effect. Figure \ref{fig:GPI} shows the wind butterfly effect in a coronagraphic image with the associated covariance map. The wind butterfly effect is strong in this image with a fractional standard deviation of 0.58, where fractional standard deviation is a measure of the azimuthal asymmetry of the point spread function on a scale from 0 to 1 \cite{Madurowicz18}. The measured effective wind speed from the associated covariance map is 24.7 m/s. This example shows promising results for the algorithm's ability to match up high effective wind speeds with prominent wind butterflies. As in the simulations, we can see two separate strong turbulent layers traversing the covariance map. Figure \ref{fig:GPI_nofly} demonstrates the lack of a strong wind butterfly with much slower turbulent layers and it isn't clear if there are multiple turbulent layers or just a single slow moving layer. The fractional standard deviation in this image is 0.22 and the calculated effective wind speed is 5.99 m/s.

\section{CONCLUSION}

Our model for calculating the effective wind speed from spatio-temporal covariance maps provides promising initial results on both simulated and real GPI WFS data. The simulated data reveals a correlation between the input effective wind speed and the effective wind speed calculated from the covariance maps. We also see promising initial results from the GPI WFS data using the same method. The next step is to apply our method to a larger sample of GPI datasets to determine if there is a clear correlation between effective wind speed and the wind butterfly effect. In addition, we aim to use a gaussian fit to identify the directions and strengths of the individual peaks instead of summing up all of the wind speed components after masking. This would allow us to more accurately calculate the effective wind speed of the atmosphere and give insight into the relationship between the wind butterfly direction and the directions of the turbulent wind layers. Further improvements can be made to the overall speed of the simulations and the covariance map calculations by rewriting the programs on a more resource efficient platform, such as CUDA. Faster calculations would make it feasible to use full resolution GPI data and maintain as much information as possible.

\acknowledgements
The GPI project has been supported by Gemini Observatory, which is operated by AURA, Inc., under a cooperative agreement with the NSF on behalf of the Gemini partnership: the NSF (USA), the National Research Council (Canada), CONICYT (Chile), the Australian Research Council (Australia), MCTI (Brazil) and MINCYT (Argentina).  This work is funded by in part by the Heising-Simons Foundation through grant 2019-1582.

\newpage
\printbibliography

\end{document}